\documentstyle[epsfig]{aipproc}

\def\ltsima{$\; \buildrel < \over \sim \;$}
\def\lsim{\lower.5ex\hbox{\ltsima}}
\def\gtsima{$\; \buildrel > \over \sim \;$}
\def\gsim{\lower.5ex\hbox{\gtsima}}

\begin{document}

{\it
Invited Review presented at the  GeV-TeV Gamma-ray Astrophysics Workshop\\
{\bf ``Towards a Major Atmospheric Cerenkov Detector VI''}\\
Snowbird, Utah - August 13-16, 1999.
}

\title{The AGILE\\ gamma-ray astronomy mission}

%\author{Steven E. Brown$^*$ and Gerard M. Steiner$^{\dagger}$}
\author{
S.~Mereghetti$^1$,    
G.~Barbiellini$^2$, 
G.~Budini$^2$,
P.~Caraveo$^1$,         
E.~Costa$^3$,        
V.~Cocco$^4$,          
G.~Di Cocco$^5$,
M.~Feroci$^3$,         
C.~Labanti$^5$,      
F.~Longo$^2$,        
E.~Morelli$^5$,       
A.~Morselli$^4$,       
A.~Pellizzoni$^6$,    
F.~Perotti$^1$,        
P.~Picozza$^4$,         
M.~Prest$^2$,
P.~Soffitta$^3$,
L.~Soli$^1$,  
M.~Tavani$^1$,  
E.~Vallazza$^2$,    
S.~Vercellone$^1$     
}

%\address{$^*$National Center for Atmospheric Research\thanks{The National
%Center for Atmospheric Research is sponsored by the National
%Science Foundation.}\\
%Boulder Colorado 80307\\
%$^{\dagger}$National Standards Institute, Boulder, Colorado 11543}

\address{
$^1$Istituto di Fisica Cosmica G.Occhialini -- CNR, Milano, Italy \\
$^2$Universit\`a di  Trieste and INFN, Trieste, Italy   \\
$^3$ Istituto di Astrofisica Spaziale -- CNR, Roma, Italy  \\
$^4$Universit\`a  ``Tor Vergata'' and INFN, Roma, Italy    \\
$^5$Istituto TESRE -- CNR, Bologna, Italy    \\
$^6$Agenzia Spaziale Italiana   \\
 }

%\lefthead{LEFT head}
%\righthead{RIGHT head}
\maketitle

\begin{abstract}
We describe the AGILE gamma-ray astronomy satellite which has recently been selected
as the first Small Scientific Mission of the Italian Space Agency.
With a launch in 2002, AGILE will provide a unique tool for high-energy astrophysics
in the 30 MeV - 50 GeV range before GLAST. 
Despite the much smaller weight and dimensions, the scientific performances of AGILE
are comparable to those of EGRET. 
%This is made possible by the silicon microstrip technology
%that allows a significant improvement in angular resolution,
\end{abstract}

\section*{Introduction}
%\[
%\widehat{a} + \widehat{ab} + \widehat{abc} + \widehat{abcd}
%\]
%
%%\show\frak
% 
%\[
%%      {\bf x}^{\bf x} \triangleq z 
%      {\bf x}^{\bf x}\triangleq{z} \tensor{T} \frak{E^E}=\frak{mc}^2
%%      {\bf x}^{\bf x}\triangleq {z} \tensor{T} \frak{E}=\frak{mc}^2
%\]
% 
%\[
%{\Bbb {NQRZ}} \qquad \because \eth\ggg\bigstar \therefore\blacktriangleright\rightsquigarrow \blacksquare
%\]
% 

The   AGILE  satellite
was  proposed in June 1997 to the   Program for Small Scientific Missions
of the Italian Space Agency (ASI). 
AGILE ({\it Astro-rivelatore Gamma a Immagini LEggero}) is a mission
devoted to  
gamma-ray (30~MeV--50~GeV) astrophysics  during the years 2002-2005
%\cite{tavani,mereghetti}.
After the initial ASI selection   in December 1997, a Phase A study
was carried out during 1998. AGILE was finally selected by ASI in 
June  1999 as the first  Small Scientific Mission  to be launched and is 
currently in  the Phase B. The launch is foreseen in early 2002. 

The AGILE  scientific payload is based on the state-of-the-art
 and reliably developed technology  of solid state silicon detectors
\cite{barbiellini1,barbiellini2,barbiellini3}.
%\cite{morselli}, \cite{barbiellini1}, \cite{barbiellini2}, \cite{barbiellini3}.
% for high-energy  photon detection and tracking.
The  instrument is  very light ($\sim 60$~kg) and  effective in
detecting and monitoring gamma-ray sources
(30~MeV--50~GeV) 
within a
large field of view.
% ($\sim 1/5$ of the whole sky).
%The baseline AGILE detector is sensitive in the energy range
%$\sim 30$~MeV--$50$~GeV,
%is characterized by the smallest ever obtained deadtime
%for gamma-ray detection   ($\loe $1~ms)
%%and by  a trigger based exclusively on silicon plane detectors.
%The instrument consists of a silicon-tungsten tracker, a cesium iodide 
%mini-calorimeter, an anticoincidence system made of segmented plastic
%scintillators, fast readout electronics and processing units.
%In constrast with old generation instruments  such as EGRET,
%AGILE does not require gas operations and/or refilling,
%and does not require high-voltages.
% Spectral information will be obtained from multiple scattering
% of created pairs in the silicon planes (for energies less than
% $\sim 500$~MeV) and by using the energy deposited in the
% tracker and a  mini-calorimeter.
The  instrument  is designed to achieve an optimal
angular resolution (source location accuracy $\sim5'-20'$ for intense sources),  
a very  large field of view
($\gsim2$~sr), and a    sensitivity comparable  to that of EGRET
for on-axis  (and substantially better for off-axis) point sources.
%%\cite{aversa, nina, barbiellini1, barbiellini2, bidoli, bocciolini, golden}.
AGILE will also carry an imaging hard X--ray detector to simultaneously
monitor in the 10-40 keV   range the sources observed in the
central part of the gamma-ray field of view.

Despite its simplicity and moderate cost,
AGILE is ideal to perform a large number of tasks \cite{tavani}:
monitoring active galactic nuclei, detecting gamma-ray bursts
with high efficiency, mapping the diffuse Galactic and extragalactic emission,
studying pulsed gamma-ray emission from radiopulsars, monitoring the many
unidentified sources and contributing to their unveiling, detecting energetic
solar flares .
Today, it is clear that successful investigations of
gamma-ray sources  rely on coordinated space and ground-based
observations.
The AGILE scientific program will be focussed on a prompt
response to gamma-ray transients and alert for follow-up
multiwavelength observations.

\begin{figure}[b!] % fig 1
\centerline{\epsfig{file=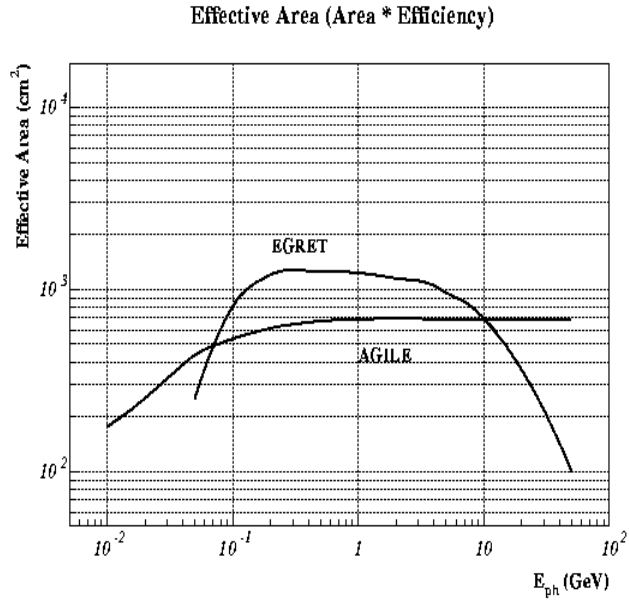,height=3.5in,width=3.5in, angle=-90}}
\vspace{10pt}
\caption{Effective area of the AGILE gamma-ray detector.}
\label{fig1}
\end{figure}

\begin{figure}[b!] % fig 2
\centerline{\epsfig{file=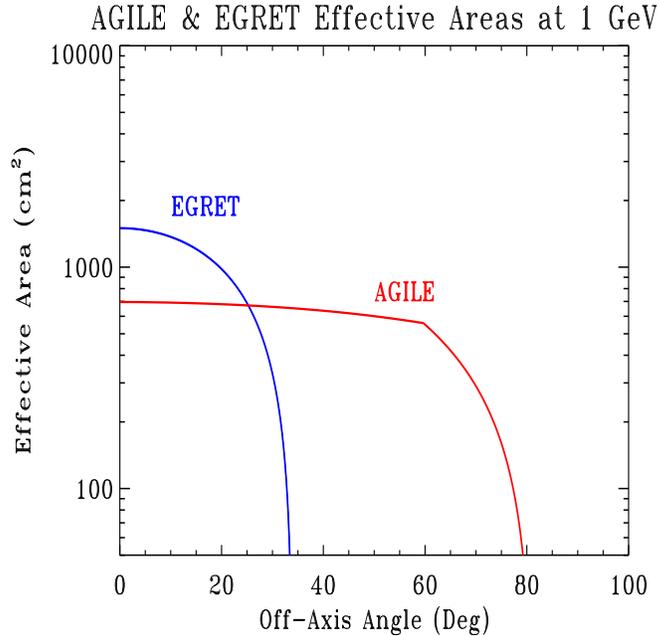,height=3.5in,width=3.5in, angle=90}}
\vspace{10pt}
\caption{Comparison of the AGILE and EGRET effective areas for
off-axis sources.}
\label{fig2}
\end{figure}

\section*{Instrument Overview }

The  AGILE scientific payload is made of three main detectors:
a silicon/tungsten Tracker, a cesium iodide Mini-calorimeter,
and a coded mask hard X-ray Imaging Detector.
These three elements are integrated in a single structure,
covered on five sides (top and lateral sides) with an
Anticoincidence for charged particles rejection.

The Tracker, consisting  of 14 planes of  silicon strip  
detectors,
will provide the unambiguous identification of incident
gamma-rays by recording the characteristic track signature
of the e$^{-}$-e$^{+}$ that result from pair creation from the incident
photons in thin layers converting material.
Each plane of the Tracker (with the exception
of the two at the bottom) 
is made of three layers: a 245 $\mu$m   thick
photon pair converter in tungsten  ($0.07 \; X_0$)
is followed by two planes  of silicon strip
detectors (thickness 410~$\mu$m) with the strips arranged 
in orthogonal directions
to provide the plane coordinates of the particle tracks.
The tungsten layer is absent in the last two planes,
since the readout trigger requires a signal in at least three consecutive
silicon planes to be activated.
The   Tracker has an  on-axis
total radiation length larger than $0.84 \; X_0$.
The resulting on-axis effective area is shown in Figure \ref{fig1}.
%corresponding to an interaction probability above 1~GeV near 35\%.
The distance between planes has been fixed to   $1.6$~cm 
on the basis of extensive  
optimizations by  Montecarlo simulations.

The fundamental unit for the silicon planes is a module of area
9.5$\times$9.5 cm$^{2}$, and  pitch (distance between strips) 
equal to 121~$\mu$m.  
Each silicon plane consists of   4x4 modules. 
The strips will be read out with  a pitch of 242~$\mu$m,
i.e. with a  floating strip every two strips,
yielding a total of $\sim$43,000 channels for the whole Tracker.
If a particle crosses a floating strip, image charges are induced on the 
two adjacent read-out  strips through capacitive coupling.
In this way, more than one strip has signal, thus 
enabling to use an interpolation algorithm  to improve the spatial
resolution. 
Special  algorithms applied off-line 
to telemetered data will allow an optimal reconstruction
of the photon incidence angle.

The  Mini-calorimeter,  
consisting of $1.5 \; X_0$ of Cesium Iodide (CsI),
will allow to determine the energy of the incident photons
imaged by the Tracker. In addition, the Mini-calorimeter 
will also be used to study gamma-ray bursts and other transient
phenomena in the $\sim$1-500 MeV energy range.
The Mini-calorimeter is located below the Tracker and consists
of  two  planes of CsI(Tl) bars. Each bar has   dimensions of  2.3$\times$1.5$\times$40 cm$^3$.
The    two planes have the bars arranged in orthogonal directions
to provide the X and Y location of the showers. 
The scintillation light   from each   bar is collected by two photodiodes
placed at both  ends.
We note that the problem of particle backscattering for
this configuration is much less severe than in the case of
EGRET, thus allowing   a relatively efficient detection of photons
up to 10~GeV.

The hard X-ray detector (Super-AGILE) is based on the 9.5x9.5 cm${^2}$  
silicon tiles that are used for  the Tracker planes. 
These will be placed on the top of the Tracker 
(above the first tungsten layer) to form an additional detection 
plane sensitive in the 10-40 keV range and used in conjunction 
with a coded mask  at a distance of about 10 cm
(below the top Anticoincidence).  
Since the silicon microstrips provide pixels only along one-dimension, 
Super-AGILE will  consist of four equal modules arranged in two pairs, 
giving monodimensional sky images along two orthogonal 
directions. 
The mask will be supported by an ultra-light structure in carbon fiber 
that will also serve as a collimator to 
reduce the Super-AGILE field of view. 
This is needed in order to limit the Super-AGILE background, which is 
dominated by the cosmic diffuse radiation.

\begin{table}[b!]
\caption{AGILE Scientific Performances}
\label{table1}
%\begin{tabular}{lrrr}
%\begin{tabular}{lccc}
\begin{tabular}{ldd}

 & Gamma-ray Detector  &      \\
\tableline
Energy Range & 30 MeV -- 50 GeV &   \\
Field of view & 2  sr  &   \\
Sensitivity at 100 MeV  &  6$\times$10$^{-9}$  ph cm$^{-2}$ s$^{-1}$ MeV$^{-1}$ &  (5$\sigma$ in 10$^{6}$ s) \\
Sensitivity at  1 GeV   &  4$\times$10$^{-11}$ ph cm$^{-2}$ s$^{-1}$ MeV$^{-1}$ &  (5$\sigma$ in 10$^{6}$ s) \\
Angular Resolution  at 1 GeV     &  36 arcmin  & (68\% containment radius) \\ 
Source Location Accuracy         &  $\sim$30 arcmin  &  for a source with S/N$\sim$10 \\
Energy Resolution           & $\Delta$E/E$\sim$1 &  at 300 MeV \\
Timing Accuracy              & 25 $\mu$s & \\
\tableline
 & Hard X--ray Detector  &      \\
\tableline
Energy Range & 10-40 keV &   \\
Field of view &  60$^{\circ}\times$60$^{\circ}$   &  Full Width at Zero Sensitivity \\
Sensitivity    &  $\sim$10 milliCrabs &  (5$\sigma$ in 1 day) \\
Angular Resolution       &  10 arcmin  &  \\ 
Source Location Accuracy   &  $\sim$2-3 arcmin  &  for a source with S/N$\sim$10 \\
Energy Resolution           & $\Delta$E$<$4 keV &    \\
Timing Accuracy              & 25 $\mu$s & \\
\end{tabular}
\end{table}

\begin{figure}[b!] % fig 3
\centerline{\epsfig{file=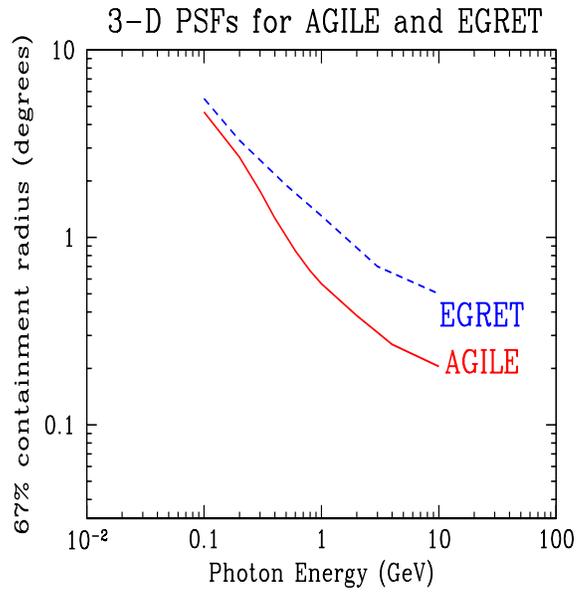 ,height=3.5in,width=3.5in, angle=-90}}
\vspace{10pt}
\caption{AGILE angular resolution.}
\label{fig3}
\end{figure}

\begin{figure}[b!] % fig 4
\centerline{\epsfig{file=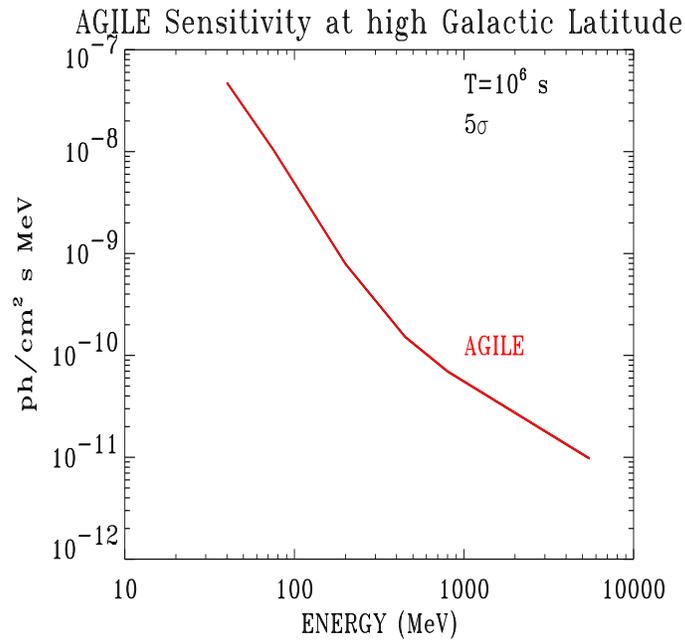,height=3.5in,width=3.5in, angle=90}}
\vspace{10pt}
\caption{Expected sensitivity for the AGILE gamma-ray instrument.}
\label{fig4}
\end{figure}

The Anticoincidence  system,  aimed at both charged 
particle background rejection and preliminary direction
reconstruction for triggered photon events,
completely surrounds the top and
lateral sides of the Super-AGILE, Tracker and Mini-calorimeter.
The top panel is a single slab of plastic scintillator with thickness
0.5 cm. The scintillation light is collected by optical fibers
glued on the four sides and directed to four photomultipier
tubes at the corners.
Each lateral face
 is segmented with three partially overlapping plastic scintillator 
layers (0.5~cm thick) connected with photomultipliers placed
at the bottom.
  % The signal from each  scintillator
% layer is collected laterally by optical fibers attached to
% photomultipliers at the bottom.
% A single plastic scintillator layer 
%(0.5~cm thick) constitutes the top-AC whose signal is read by four
%light photomultipliers  placed inside the AC system  and supported
%by the four corners of the structure frame.

\section*{Performances}

The expected scientific performances of AGILE are summarized in 
Table \ref{table1}. One of the main characteristics of the AGILE gamma-ray
detector is the very large field of view. As shown in Figure \ref{fig2}
the AGILE effective area remains almost constant for large off-axis angles.
This will allow to simultaneously monitor a large number of sources in a single 
pointing and will also result, at the end of the mission, in a large exposure
factor for each region of the sky \cite{mereghetti}.

The great sky exposure factor, coupled with the good angular resolution
(see Figure \ref{fig3}), will allow a detailed study of the diffuse
Galactic and Extra-Galactic emission and to better locate the unidentified EGRET 
sources. 

Another important characteristic of the AGILE Tracker is the small dead time.
This will be crucial in the study of the high energy emission of gamma-ray bursts,
which are expected in the field of view at a rate of $\sim$5-10 per year, 
based on the EGRET results.

The extension in the hard X--ray range of a gamma-ray mission, made possible
by the Super-AGILE detector, is an innovative concept that will allow the 
study of correlated variability for sources of different classes, ranging from 
active galactic nuclei and blazars to unidentified galactic transients.

\section*{The Mission}

The AGILE instrument will be carried in an equatorial 
(inclination $<$5$^{\circ}$), circular orbit (altitude $\sim$550 
km), by a spacecraft of the MITA class, which is currently being 
developed by Gavazzi Space as prime contractor. 
The total mass will be of the order of 180-200 kg,  
including  $\sim$60 kg of scientific  payload. 

The satellite will point 
with a 3-axis attitude stabilization with an accuracy of 
the order of 1$^{\circ}$. An attitude reconstruction at the   
arcmin level will be obtained a posteriori, by means of star sensors.
A typical AGILE pointing will last 2-3 weeks.  
To maximise the observing efficiency, we are investigating  the 
possibility of slewing to secondary pointing directions 
during the fraction of the orbit in which the primary target 
direction is occulted by the Earth. 
AGILE will also have the possibility to quickly repoint (within 1 day) 
in order to react 
to interesting targets of opportunity. 

The operations center
will have the primary responsibility of satellite 
operations and communications using the ASI ground base at Malindi (Kenya). 
The equatorial orbit will allow a single 
contact per orbit at a downlink rate of 500 kbit s$^{-1}$. 
 
An AGILE data center will be devoted to the monitoring of the 
instrument scientific performance, to the quick look analysis, 
and to  the processing, distribution and archival of 
scientific data products and the associated calibration data.

\section*{Conclusions}

AGILE will provide crucial  information  complementary to the
many lower energy detectors that will be operational
during the first decade of the new Millenium
(INTEGRAL, XMM, AXAF, ASTRO-E, SPECTRUM-X, and others).
% probably BSAX).
No other gamma-ray mission  dedicated to the energy band above $\sim 30$~MeV
is planned before GLAST (which will most likely be operational 
after   2005-2006). 
AGILE might then  be an ideal `bridge' between old and new generation
gamma-ray missions  with an innovative design and efficient scientific
management.

\end{document}